\documentclass[twocolumn]{jpsj3}

\title{
Antisymmetric Spin-Orbit Coupling Effect \\
on Kondo-Induced Electric Polarization \\
in a Triangular Triple Quantum Dot
}

\author{Mikito Koga$^1$, Masashige Matsumoto$^2$, and Hiroaki Kusunose$^3$}

\inst{
$^1$Department of Physics, Faculty of Education, Shizuoka University, Shizuoka
422--8529, Japan \\
$^2$Department of Physics, Faculty of Science, Shizuoka University, Shizuoka
422--8529, Japan \\
$^3$Department of Physics, Meiji University, Kawasaki 214-8571, Japan
}

\abst{
We study the local antisymmetric spin-orbit (ASO) coupling effect on spin, orbital, and charge
degrees of  freedom for the Kondo effect in a triangular triple quantum dot (TTQD).
Here, one of the three QDs is coupled to a metallic lead through electron tunneling, and
a local electric polarization is induced by the Kondo effect.
The ASO interaction is introduced in the other two coupled QDs on the opposite side of the lead.
Generally, the ASO coupling effect is very weak and not easily detectable, but it essentially
causes spin and charge reconfigurations in the TTQD through the Kondo effect.
Using an extended Anderson model for the TTQD Kondo system, we elucidate that the ASO
coupling gives rise to a considerable reduction of the emergent electric polarization, as a consequence
of the parity mixing of molecular orbitals in the triangular loop as well as the spin-up and
spin-down coupling of local electrons.
The latter leads to a local diamagnetic susceptibility owing to the ASO coupled spins.
We also show that the Kondo-induced electric polarization can be controlled by the ASO coupling
as well as by the magnetic flux penetrating through the TTQD.
}

\begin{document}

\maketitle

\newcommand{\ds}{\displaystyle}
\newcommand{\bS}{{\mbox{\boldmath$S$}}}
\newcommand{\bk}{{\mbox{\boldmath$k$}}}
\newcommand{\bskp}{{\mbox{\scriptsize\boldmath $k$}}}
\newcommand{\skp}{{\mbox{\scriptsize $k$}}}
\newcommand{\bsk}{\bskp}
\newcommand{\ri}{{\rm i}}
\newcommand{\bsig}{{\mbox{\boldmath$\sigma$}}}
\newcommand{\bl}{{\mbox{\boldmath$l$}}}
\newcommand{\bs}{{\mbox{\boldmath$s$}}}

\section{Introduction}
The Kondo effect is a strongly correlated quantum phenomenon originating from a local spin
coupled to conduction electrons antiferromagnetically owing to the strong repulsion between
localized electrons in atomic orbitals.
\cite{Hewson93}
The low-temperature physics is well explained by the Fermi liquid picture, in which the local
spin is completely screened by the conduction electrons through the formation of a spin singlet called the Kondo singlet.
The Kondo physics has been extensively studied in a wide range of fields from $d$- or
$f$-electron impurities embedded in bulk materials
\cite{Cox98,Kondo05}
to artificial atomic systems with metallic lead contacts such as
quantum dot (QD) or magnetic molecular devices.
\cite{Wiel02,JarilloHerrero05,Hanson07,Roch08,Parks10,Laird15}
In particular, much attention has been paid to rich Kondo phenomena associated with
coupled spin clusters or fabricated quantum devices, for instance, the transport properties
between leads through a double
\cite{Wiel02,Hanson07,Izumida97,Lopez02,Tanaka05,Wang11,Okazaki11,Amasha13}
or triple QD.
\cite{Schroer07,Yamahata09,Amaha12,Amaha13,Busl13}
\par

As a different context, several theoretical studies have recently investigated the Kondo effect for a
single impurity in two-dimensional electron gas with the Rashba spin-orbit (SO) coupling.
\cite{Malecki07,Zitko11,Yanagisawa12,Zarea12,Isaev12,Wong16}
The Rashba coupling is one of the antisymmetric spin-orbit (ASO) interactions and has
attracted much attention from numerous researchers of bulk systems, such as the spintronics in
semiconductors,
\cite{Zutic04,Ando17}
topological insulators,
\cite{Qi11,Ando13}
and non-centrosymmetric superconductors.
\cite{Sato16}
For the Rashba Kondo system, it has been found that the low-temperature physics meets the
Fermi liquid picture as described by the conventional Kondo model, although the Kondo
temperature can be considerably increased in certain situations for a relatively strong coupling of the Rashba SO
interaction.
\cite{Wong16}
All the above studies devoted themselves to the SO coupling effects on itinerant or band
electrons.
\par

On the nanoscale, an ASO coupling also arises in coupled atoms with different-parity
orbitals when the inversion symmetry is absent.
Let us consider the transfer of tight-binding electrons between two atoms through the overlap of
localized orbitals.
Owing to the absence of the inversion symmetry, a local electric field gives rise to the mixing of
intraatomic even- and odd-parity orbitals, leading to an effective spin and orbital exchange
interaction on the electron hopping between the two sites.
\cite{Yanase08}
The orbital exchange occurs between symmetric-bonding and antisymmetric-bonding states of
the coupled atoms, which resembles the antisymmetric $\bk$ dependence of the ASO interactions
in bulk systems, where $\bk$ is the wave vector of an itinerant electron.
It is expected that this ASO interaction plays an important role in the parity mixing of
degenerate molecular orbitals in coupled magnetic impurities such as a triangular triple quantum
dot (TTQD).
\par

For the last decade, the TTQD has been experimentally realized in AlGaAs/GaAs heterostructures
\cite{Vidan04,Gaudreau06,Rogge08,Amaha09,Seo13}
and self-assembled InAs systems.
\cite{Amaha08}
The recent development of a fabrication technique has stimulated theoretical investigations of
various TTQD Kondo systems because of the high potentiality for versatile quantum devices.
\cite{Kuzmenko06,Oguri07,Wang07,Zitko08,Wang08,Mitchell09,Vernek09,Oguri11,Mitchell13,
Tooski14,Oguri15,Xiong16,Tooski16}
The tunable parameters in the TTQD are interdot electron-hopping matrix elements and
intradot orbital energy levels used as a standard theoretical setup.
The electron occupation number at each dot changes with the depth of the orbital energy relative
to the Coulomb coupling.
In particular, in a half-filled case, the three-site spins are coupled antiferromagnetically owing to
the superexchange process between two sites with opposite spins.
This spin frustration competes with the Kondo effect and leads to a quantum transition associated 
with the stabilization or destabilization of Kondo singlet formation, which was investigated for 
several configurations of dot-lead contacts.
\cite{Zitko08,Wang08,Mitchell09,Mitchell13,Tooski14,Xiong16}
The loop structure of the TTQD also gives rise to interference effects such as the Aharonov--Bohm
(AB) effect,
\cite{Izumida97,Akera93,Bruder96,Hofstetter01,Tanaka06,Komijani13}
where a magnetic flux penetrating through the loop affects the molecular orbitals of
the TTQD and modifies the Kondo behavior in the absence of a magnetic field.
\cite{Kuzmenko06,Wang08}
However, most of the theoretical studies have mainly focused on the conductance through the
TTQD as an observable and controllable physical quantity.
\par

\begin{figure}
\begin{center}
\includegraphics[width=7cm,clip]{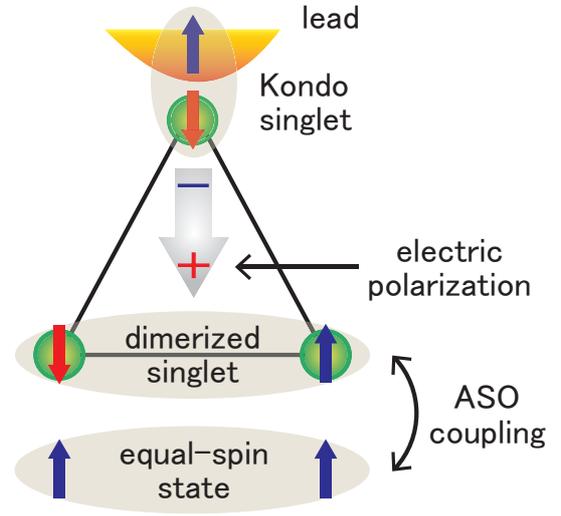}
\end{center}
\caption{
(Color online)
Illustration of the TTQD Kondo system.
When the ASO coupling ($\lambda_{\rm A} = \gamma t$) is absent, the Kondo singlet is realized
at the top site (labeled $a$ in the model), which is accompanied by a dimerized spin singlet at the
bottom ($b$-$c$ bond in the model).
This spin reconfiguration induces charge redistribution in the TTQD and electric
polarization (KIEP) at low temperatures (represented by the arrow at the center of the triangle).
The local ASO coupling introduced in the $b$-$c$ bond causes an effective spin and orbital
exchange as well as parity mixing of the $E_\pm$ molecular orbitals, which leads to a
considerable reduction of the electric polarization for a small Kondo coupling strength.
Simultaneously, equal-spin states with $S_z = \pm 1$ are mixed with the $b$-$c$ singlet in the
ground state and this mixing weight is increased by the ASO coupling.
}
\label{fig:aso}
\end{figure}

As mentioned above, the parity mixing of molecular orbitals is another important feature of
the closed loop of the TTQD.
\cite{Koga12,Koga16}
The charge number at each site is closely correlated with the three-site spin configuration, and
the deviation of charge depends on the third order of the intersite electron hopping parameter.
\cite{Bulaevskii08}
In this paper, we consider the local ASO coupling as a source of orbital-parity mixing and
demonstrate how the parity mixing affects the electric polarization induced by
the Kondo effect, as illustrated in Fig.~\ref{fig:aso} (the details are described in Sect.~3).
The ASO interaction can be realized in the absence of the inversion symmetry related to the
two-dimensionality of triangular arrangements of atoms or QDs.
Generally, the orbital-parity mixing is not detectable since the bare value of the local ASO
coupling is rather small.
However, this ASO coupling affects the Kondo screening process in an essential way.
When the ASO interaction is absent, the Kondo-induced electric polarization (KIEP) is generated
by one of two different-parity orbitals of the TTQD strongly coupled to lead electrons in the
Kondo singlet formation.
For a finite ASO coupling, the other orbital also participates in the Kondo singlet and
gives rise to a considerable reduction of the emergent electric polarization.
Since these orbitals are coupled with spins, the local spin susceptibility exhibits a unique ASO
coupling dependence.
This is different from another orbital-mixing effect on the KIEP
associated with the AB effect, which has been reported in our recent paper.
\cite{Koga14}
Here, we also compare the control of the electric polarization via the ASO coupling and magnetic
flux to elucidate a common feature of the parity mixing of molecular orbitals in the TTQD.
\par

This paper is organized as follows.
In Sect.~2, the local ASO interaction is described in the case of coupled QD sites, and an essential
role of the ASO coupling is demonstrated by a two-site Hubbard model.
In Sect.~3, an extended Anderson model for the TTQD is given, which is used to investigate the
interplay between the ASO coupling and Kondo effects on the emergent electric polarization, i.e.,
KIEP.
In Sect.~4, the KIEP and local spin susceptibility are calculated as a function of a parity-mixing
parameter corresponding to the ASO coupling using Wilson's numerical renormalization group
(NRG) method.
\cite{Wilson75,Krishnamurthy80,Bulla08}
For comparison, a similar parity-mixing effect is also shown in the presence of a magnetic flux
through the triangular loop as a different context.
The summary and concluding remarks are given in the last section.
In Appendix~A, a derivation of the local ASO interaction is shown by analogy with atomic states.
In Appendix~B, the lowest-energy states of a TTQD are described in a half-filled case of a
three-site Hubbard model, which is used to explain the ASO coupling effect on KIEP in Sect.~4.
Appendix~C gives a brief review of the magnetic flux effect on an isolated TTQD.
\par

\section{Local Antisymmetric Spin-Orbit Interaction}
First, we consider an ASO interaction between two sites labeled 1 and 2 as expressed by the
following Hamiltonian:
\begin{align}
H_{\rm so}(1,2) = \lambda_{\rm A}
( d_{1 \uparrow}^\dagger d_{2 \downarrow} - d_{2 \uparrow}^\dagger d_{1 \downarrow}
 - d_{1 \downarrow}^\dagger d_{2 \uparrow} + d_{2 \downarrow}^\dagger d_{1 \uparrow} ),
\label{eqn:Hso1}
\end{align}
where $d_{i \sigma}^\dagger$ ($d_{i \sigma}$) is a creation (annihilation) operator for a localized
electron at the $i$th site with spin $\sigma$ ($= \uparrow, \downarrow$) and
$\lambda_{\rm A}$ is a coupling constant.
This local interaction can be derived by onsite and intersite hybridizations between different-parity
atomic orbitals in addition to an onsite SO interaction between degenerate orbitals.
\cite{Yanase11}
As described in Appendix~A, it becomes relevant for multiorbital electron systems in the absence
of the inversion symmetry, which resembles the ASO interactions of band electrons in
two-dimensional systems.
For simplicity, we have assumed that the two sites are aligned in the $x$ direction and the
intersite orbital-parity mixing is owing to a deviation from the inversion symmetry in the $z$
direction.
In the two-site electron system, we define the even-parity (symmetric-bonding) and odd-parity
(antisymmetric-bonding) orbital electrons with new
operators as
\begin{align}
d_{\rm e \sigma}^\dagger = \frac{ d_{1 \sigma}^\dagger + d_{2 \sigma}^\dagger }{\sqrt{2}},~~
d_{\rm o \sigma}^\dagger = \frac{ d_{1 \sigma}^\dagger - d_{2 \sigma}^\dagger }{\sqrt{2}},
\label{eqn:eo}
\end{align}
respectively, since they are transformed as
$d_{{\rm e} \sigma}^\dagger \rightarrow d_{{\rm e} \sigma}^\dagger$ and
$d_{{\rm o} \sigma}^\dagger \rightarrow - d_{{\rm o} \sigma}^\dagger$ with respect to the
interchange of the site indices $1 \leftrightarrow 2$.
Using the orbital basis in Eq.~(\ref{eqn:eo}), Eq.~(\ref{eqn:Hso1}) is rewritten as
\begin{align}
H_{\rm so} = \lambda_{\rm A}
( - d_{{\rm e} \uparrow}^\dagger d_{{\rm o} \downarrow}
 + d_{{\rm o} \uparrow}^\dagger d_{{\rm e} \downarrow}
 + d_{{\rm e} \downarrow}^\dagger d_{{\rm o} \uparrow}
 - d_{{\rm o} \downarrow}^\dagger d_{{\rm e} \uparrow} ).
\label{eqn:Hso2}
\end{align}
Using pseudospin components $\tau = \uparrow$ and $\tau = \downarrow$ for the orbital indices
'e' and 'o', respectively, the spin and orbital exchange in Eq.~(\ref{eqn:Hso2}) is simplified to
\begin{align}
H_{\rm so} = \lambda_A
 \sum_{\tau \tau' \sigma \sigma'} d_{\tau \sigma}^\dagger \left( \tau_y \right)_{\tau \tau'}
 \left( \sigma_y \right)_{\sigma \sigma'} d_{\tau' \sigma'},
\end{align}
where $\sigma_y$ and $\tau_y$ are the $y$ components of the Pauli matrices.
The $\tau_y \sigma_y$ interaction does not conserve the $z$ component of composite spins
of the two-site electrons.
Instead, the spin and orbitally coupled states are classified by
\begin{align}
\eta_z = \prod_{n=1}^N (\tau_z \sigma_z)_n = \pm 1
\label{eqn:eta}
\end{align}
for $N$ electrons, and each $(\tau_z \sigma_z)_n$ takes a value of either $+1$ or $-1$ for
the $n$th electron state with $\tau, \sigma = \uparrow, \downarrow$.
\cite{Note1}
We categorize the vacuum state into $\eta_z = +1$.
\par

Let us demonstrate how different spin states of electrons are coupled via
the ASO interaction using the two-site Hubbard model
\begin{align}
& H_{\rm dimer} = U ( n_{1 \uparrow} n_{1 \downarrow} + n_{2 \uparrow} n_{2 \downarrow} )
\nonumber \\
&~~~~~~
 - t \sum_\sigma ( d_{1 \sigma}^\dagger d_{2 \sigma} + d_{2 \sigma}^\dagger d_{1 \sigma} )
 + H_{\rm so} (1,2),
\label{eqn:Hdimer}
\end{align}
where $n_{i \sigma}$ ($\equiv d_{i \sigma}^\dagger d_{i \sigma}$) represents the number of
$i$th-site electrons with $\sigma$, $U$ ($> 0$) is the Coulomb coupling, and $t$ ($> 0$) is
the electron-hopping matrix element between the two sites.
At half-filling, we consider the lowest-energy state for $t / U \ll 1$ dominated by the following
spin-singlet state:
\begin{align}
| \phi_{\rm S} \rangle = \frac{1}{\sqrt{2}}
 ( d_{1 \uparrow}^\dagger d_{2 \downarrow}^\dagger
 - d_{1 \downarrow}^\dagger d_{2 \uparrow}^\dagger ) | 0 \rangle,
\label{eqn:phiS}
\end{align}
where $| 0 \rangle$ represents a vacuum state.
This is combined with the doubly occupied state
\begin{align}
| \phi_{\rm D} \rangle = \frac{1}{\sqrt{2}}
 ( d_{1 \uparrow}^\dagger d_{1 \downarrow}^\dagger
 + d_{2 \uparrow}^\dagger d_{2 \downarrow}^\dagger ) | 0 \rangle
\end{align}
through the transition matrix element
$\langle \phi_{\rm D} | H_{\rm dimer} | \phi_{\rm S} \rangle = - 2 t$.
Both states are categorized into the $\eta_z = -1$ type.
Indeed, their orbital-basis wave functions are given as
\begin{align}
& | \phi_{\rm S} \rangle = \frac{1}{\sqrt{2}}
 ( d_{{\rm e} \uparrow}^\dagger d_{{\rm e} \downarrow}^\dagger
 - d_{{\rm o} \uparrow}^\dagger d_{{\rm o} \downarrow}^\dagger ) | 0 \rangle,
\nonumber \\
& | \phi_{\rm D} \rangle = \frac{1}{\sqrt{2}}
 ( d_{{\rm e} \uparrow}^\dagger d_{{\rm e} \downarrow}^\dagger
 + d_{{\rm o} \uparrow}^\dagger d_{{\rm o} \downarrow}^\dagger ) | 0 \rangle.
\end{align}
One can find that the doubly occupied (e, $\uparrow$) and (e, $\downarrow$) electron states
lead to
$\eta_z = (\tau_z \sigma_z)_{{\rm e} \uparrow} \times (\tau_z \sigma_z)_{{\rm e} \downarrow}
= (+ 1) \times (-1) = -1$.
The same algebra is also applied to (o, $\downarrow$) with $\tau_z \sigma_z = +1$ and
(o, $\uparrow$) with $\tau_z \sigma_z = -1$.
When $\lambda_{\rm A}$ is finite, $| \phi_{\rm D} \rangle$ is coupled to an equal-spin state
($y$ component of a spin-triplet state) with $\eta_z = -1$,
\begin{align}
& -i | \phi_{{\rm T}_y} \rangle = \frac{1}{\sqrt{2}}
 ( d_{1 \uparrow}^\dagger d_{2 \uparrow}^\dagger
 + d_{1 \downarrow}^\dagger d_{2 \downarrow}^\dagger ) | 0 \rangle
\nonumber \\
&~~~~~~~
 = \frac{1}{\sqrt{2}}
 ( d_{{\rm o} \uparrow}^\dagger d_{{\rm e} \uparrow}^\dagger
 + d_{{\rm o} \downarrow}^\dagger d_{{\rm e} \downarrow}^\dagger ) | 0 \rangle,
\end{align}
through
$i \langle \phi_{{\rm T}_y} | H_{\rm dimer} | \phi_{\rm D} \rangle = - \lambda_{\rm A}$.
On the basis of the above three states, the diagonal matrix elements of $H_{\rm dimer}$ are as
follows:
\begin{align}
& \langle  \phi_{\rm S} | H_{\rm dimer} | \phi_{\rm S} \rangle
 = \langle \phi_{{\rm T}_y} | H_{\rm dimer} | \phi_{{\rm T}_y} \rangle = 0,
\nonumber \\
& \langle \phi_{\rm D} | H_{\rm dimer} | \phi_{\rm D} \rangle = U.
\end{align}
After diagonalizing the $3 \times 3$ matrix of $H_{\rm dimer}$, we obtain the lowest-energy state
\begin{align}
& | \psi_{\rm g} \rangle \simeq \frac{1 - \delta}{\sqrt{1 + \gamma^2}} | \phi_{\rm S} \rangle
 + 2 \bar{t} \sqrt{1 + \gamma^2} | \phi_{\rm D} \rangle
\nonumber \\
&~~~~~~
 + \frac{\gamma (1 - \delta)}{\sqrt{1 + \gamma^2}} ( -i ) | \phi_{{\rm T}_y} \rangle,
\label{eqn:psiSDT}
\end{align}
which is solved up to the second order of $\bar{t}$ ($\equiv t / U$), where $\gamma$
($\equiv \lambda_{\rm A} / t$) is used and $\delta = 2 \bar{t}^2 (1 + \gamma^2)$.
Thus, the different spin states $| \phi_{\rm S} \rangle$ with $S_z = 0$ and
$-i | \phi_{{\rm T}_y} \rangle$ with $S_z = \pm 1$ are coupled to each other by the ASO interaction
through $| \phi_{\rm D} \rangle$.
As a consequence, the vector spin chirality
$\langle \psi_{\rm g} | (\bS_1 \times \bS_2)_y | \psi_{\rm g} \rangle
\simeq \gamma / (1 + \gamma^2)$ is generated, where $\bS_i$ ($i = 1, 2$) is the spin operator
($\bS \equiv \bsig / 2$) at the $i$th site.
\cite{Note2}
This is closely related to the emergent electric polarization due to the mixing of different-parity
orbitals in the absence of the inversion symmetry.
\cite{Matsumoto17}
\par

For the later calculation, we also show the roles of the ASO coupling in a half-filled
three-site Hubbard model for the strong Coulomb coupling $\bar{t} \ll 1$ in Appendix~B.
As in the above two-site case, the ASO coupled spins generate a vector spin chirality
perpendicular to the axis along which the two spins are aligned.

\section{Model}
To elucidate the role of the ASO interaction in strongly correlated electron systems, we study
the Kondo effect in the TTQD, where one of the three sites is connected to a metallic lead through
electron tunneling.
In such an artificial molecule, spin and charge reconfigurations are generated by the Kondo effect.
This nanoscale magnetoelectric effect can be controlled by an intersite ASO interaction in the spin
cluster if the ASO coupling strength is tunable.
For instance, the coupling constant $\lambda_A$ in Eq.~(\ref{eqn:Hso1}), namely, $\gamma$ in Eq.~(\ref{eqn:gamma}), depends on the onsite and intersite orbital-parity mixing $V_z$ and
$t_{sp}$, respectively.
The former is associated with the absence of the inversion symmetry that induces a local electric field in the $z$ direction,
\cite{Yanase11}
which can be fine-tuned by gate-voltage control in nanoscale devices.
Here, we introduce the ASO interaction in only one bond of the TTQD on the opposite side of the
lead, which breaks the equivalency of the three QDs.
However, point-group representations of the triangular symmetry are still useful for classifying
local electron states of the TTQD.
As a major advantage of this simplification, we can demonstrate more clearly the interplay
between the ASO coupling and Kondo effects through the different-parity mixing of molecular
orbitals in the TTQD, since the local ASO interaction does not affect the Kondo effect directly and
can be regarded as a moderate perturbation.
\par

We investigate the above ASO coupling effect using an extended Anderson model Hamiltonian
that consists of three terms,
$H=H_{\rm d} + H_{\rm l} + H_{\rm l-d}$, for the TTQD, the kinetic energy of lead electrons, and
the electron tunneling, respectively.
The first term is represented by the following Hubbard-type model frequently used for the TTQD:
\begin{align}
& H_{\rm d} = -t \sum_{i \ne j} \sum_\sigma d_{i \sigma}^\dagger d_{j \sigma}
 + \sum_i \left( \varepsilon_{\rm d} + \frac{U}{2} \right) n_i
\nonumber \\
&~~~~~~
 +\frac{U}{2} \sum_i (n_i - 1)^2 +H_{\rm so} (b,c)~~(i,j = a,b,c).
\label{eqn:Hd}
\end{align}
Here, the three sites are identical, except for the presence of the ASO interaction in
Eq.~(\ref{eqn:Hso1}) between the $b$- and $c$-sites.
The onsite electron number $n_i$ ($\equiv n_{i \uparrow} + n_{i \downarrow}$) depends on the
depth of the local orbital energy $\varepsilon_{\rm d}$ ($<0$).
Here, we choose the symmetric condition $\varepsilon_{\rm d} = - U / 2$ that favors
single-electron occupation at each site and generates a local spin.
In the lead-electron term
\begin{align}
H_{\rm l} = \sum_{\bsk \sigma} \varepsilon_\bsk c_{\bsk \sigma}^\dagger c_{\bsk \sigma},
\end{align}
the creation (annihilation) of conduction electrons is represented by $c_\bsk^\dagger$ ($c_\bsk$)
with the wave number $\bk$ and spin $\sigma$.
We consider the electron tunneling between one of the three sites ($a$-site) and the lead
(hybridization between the $a$-site orbital and a conduction band) as
\begin{align}
H_{\rm l-d} = \sum_{\bsk \sigma} ( v_\bsk d_{a \sigma}^\dagger c_{\bsk \sigma} + {\rm h. c.} ),
\end{align}
where $v_\bsk$ is related to the level broadening of $\Gamma \equiv \pi \rho | v_\bsk |^2$
($\rho$ is the density of electron states at the Fermi energy), which is considered as a constant.
As mentioned in the previous section, the total $S_z$ is not conserved by the ASO interaction.
Instead, the even or odd symmetry in terms of $\eta_z = \pm 1$ in Eq.~(\ref{eqn:eta}) is useful for
classification of the spin and orbitally coupled states.
Since the lead is only connected to the $a$-site, the relevant conduction electrons belong to
the same $\tau_z \sigma_z$ types of $a$-site electrons, namely,
\begin{align}
c_{\bsk \uparrow}^\dagger \rightarrow \tau_z \sigma_z = +1,~~
c_{\bsk \downarrow}^\dagger \rightarrow \tau_z \sigma_z = -1.
\end{align}
\par

In the present study, we calculate the electric polarization in the triangular cluster defined as
\begin{align}
& \delta n = \frac{1}{3} ( 2 \langle n_a \rangle - \langle n_b \rangle - \langle n_c \rangle )
\nonumber \\
&~~~~
 = \frac{1}{3} ( 2 \langle n_a \rangle - \langle n_{\rm e} \rangle - \langle n_{\rm o} \rangle ),
\label{eqn:dn}
\end{align}
where $\langle \cdots \rangle$ represents the expectation value.
The electron number in the even-parity orbital ($n_{\rm e}$) or odd-parity orbital ($n_{\rm o}$) is
defined as $n_\tau = \sum_\sigma d_{\tau \sigma}^\dagger d_{\tau \sigma}$ ($\tau$ = e,o), where
$d_{{\rm e} \sigma}^\dagger = ( d_{b,\sigma}^\dagger + d_{c,\sigma}^\dagger ) / \sqrt{2}$ and
$d_{{\rm o} \sigma}^\dagger = ( d_{b,\sigma}^\dagger - d_{c,\sigma}^\dagger ) / \sqrt{2}$.
For the isolated TTQD ($\Gamma = 0$), each of the lowest-energy states in
Eq.~(\ref{eqn:psiE}) gives the opposite expectation value to the corresponding electric polarization
operator represented by
$\delta \hat{n} = (2 n_a - n_b - n_c) / 3$ as
\cite{Koga12,Bulaevskii08}
\begin{align}
\langle \psi_\sigma^{E_+} | \delta \hat{n} |  \psi_\sigma^{E_+} \rangle
= - \langle \psi_\sigma^{E_-} | \delta \hat{n} |  \psi_\sigma^{E_-} \rangle
= 12 \left( \frac{t}{U} \right)^3.
\label{eqn:dnE}
\end{align}
The $E_\pm$ degeneracy causes cancellation of the net electric polarization and leads to
$\delta n = 0$.
Thus, the appearance of a finite $\delta n$ is associated with the energy difference between the
$E_\pm$ states.
For a finite $\gamma$, the opposite-spin state $| \psi_{- \sigma}^{E_-} \rangle$ is coupled to
$| \psi_\sigma^{E_+} \rangle$.
When the $E_+$ state is considerably lowered in energy by the Kondo effect, the $A_-$ states
with the same $\eta_z$ also participate in the lowest-energy state together with the $E_-$ state as
in Eq.~(\ref{eqn:psi4}).
A schematic picture of the spin configuration is given in Fig.~\ref{fig:aso}.
It is expected that this ASO interaction affects the emergence of $\delta n$ and gives rise to a spin
reconfiguration in the TTQD, the details of which are shown by the NRG analysis in the
next section.
\par

In the NRG calculation, the logarithmic discretization parameter $\Lambda = 3$ is used for the
conduction band, the half width of which is chosen as the energy unit.
\cite{Wilson75,Krishnamurthy80}
The physical temperature $T$ is also measured in terms of this unit. 
Here, the NRG results are shown for a strong Coulomb coupling that is fixed at $U = 0.9$.
At each renormalization step, about 2000 low-lying states are retained.
All wave functions are classified using $\eta_z = \pm 1$ in Eq.~(\ref{eqn:eta}) for arbitrary electron
numbers $N$ in the lead plus the TTQD.

\section{Results}
\subsection{Local electric polarization and magnetic susceptibility}
In the previous study, we showed the KIEP in the absence of the ASO interaction ($\gamma = 0$).
\cite{Koga12}
As the temperature $T$ is decreased, $\delta n$ in Eq.~(\ref{eqn:dn}) exhibits an abrupt increase
and reaches a constant value $\delta n^*$ at low $T$.
The saturation value $\delta n^*$ mainly depends on $t / U$ and monotonically increases
with decreasing the Kondo coupling strength $\Gamma / U$.
As $\Gamma / U$ approaches zero, $\delta n^*$ increases up to a maximum of $12 (t / U)^3$ for
$t / U \ll 1$, which corresponds to
$\langle \psi_\sigma^{E_+} | \delta \hat{n} | \psi_\sigma^{E_+} \rangle$ at
$\Gamma \rightarrow 0$ in Eq.~(\ref{eqn:dnE}), indicating that $| \psi_\sigma^{E_+} \rangle$ is
stabilized by the Kondo effect.
\par

\begin{figure}
\begin{center}
\includegraphics[width=7cm,clip]{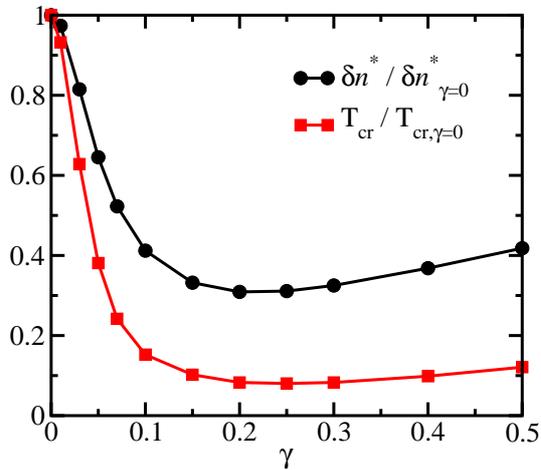}
\end{center}
\caption{
(Color online)
Saturation value $\delta n^*$ of the emergent electric polarization at low temperatures and
crossover temperature $T_{\rm cr}$ evaluated by the local magnetic susceptibility $\chi$.
The data normalized by each $\gamma = 0$ value are plotted as a function of $\gamma$
associated with the ASO coupling strength $\gamma t$ for $t / U = 0.12$ and
$\Gamma / U = 0.0473$.
}
\label{fig:nxtcr}
\end{figure}

\begin{figure}
\begin{center}
\includegraphics[width=7cm,clip]{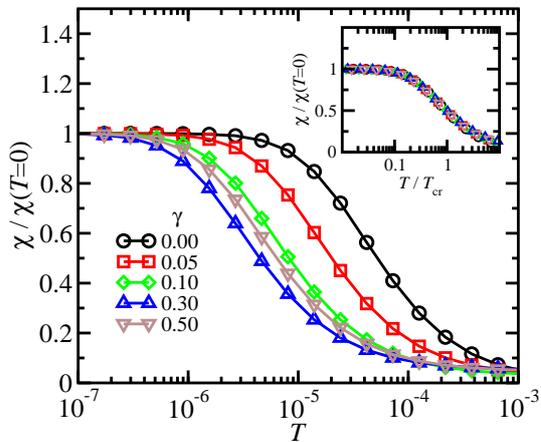}
\end{center}
\caption{
(Color online)
Temperature $T$ dependence of local magnetic susceptibility of TTQD for various $\gamma$
values of the ASO coupling.
The data are normalized by $\chi (T=0)$.
Here, $t / U$ and $\Gamma / U$ are fixed at the same values as in Fig.~\ref{fig:nxtcr}.
Inset: Universal $\chi / \chi (T=0)$ as a function of $T / T_{\rm cr}$.
Here, $T_{\rm cr}$ is determined as the temperature for $\chi / \chi (T=0) = 1/2$.
}
\label{fig:chi}
\end{figure}

Figure~\ref{fig:nxtcr} shows the ASO coupling effect on $\delta n^*$ and the crossover temperature
$T_{\rm cr}$.
For $t = 0$, the latter corresponds to the conventional Kondo temperature in the Kondo effect due
to a single local spin.
For a finite $t$, the ASO coupling $\gamma$ (in the unit of $t$) gives rise to the spin and $E_\pm$
orbital mixing, leading to a reduction of $T_{\rm cr}$.
The $\gamma$ dependence of $T_{\rm cr}$ can be evaluated by a universal fitting
of the local magnetic susceptibility $\chi (T)$ of the TTQD for various $\gamma$ values.
This $\chi$ is calculated under the assumption that the applied magnetic field is perpendicular to
the TTQD, i.e., parallel to the threefold axis of the triangular symmetry.
The result is that the data of $\chi / \chi (T=0)$ in Fig.~\ref{fig:chi} are unified into a universal
function of the rescaled temperature $T / T_{\rm cr}$.
This indicates that the Kondo singlet is also realized at $T < T_{\rm cr}$ for a finite ASO coupling.
\par

\begin{figure}
\begin{center}
\includegraphics[width=7cm,clip]{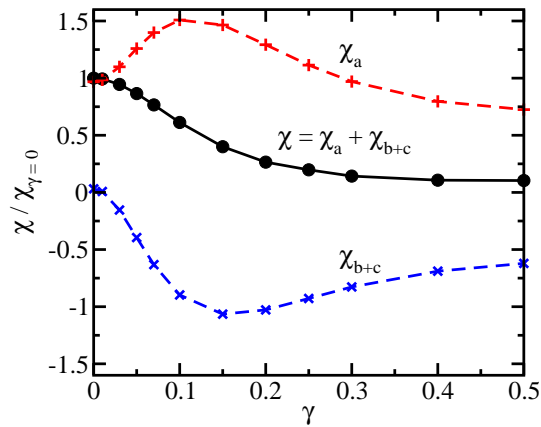}
\end{center}
\caption{
(Color online)
Local magnetic susceptibility of TTQD at the zero temperature $\chi = \chi_a + \chi_{b+c}$,
where $\chi_a$ and $\chi_{b+c}$ are the onsite susceptibilities for the $a$-site and
the $b$- plus $c$-sites, respectively.
The data are normalized by the $\gamma = 0$ value $\chi_{\gamma = 0}$ of the total $\chi$
and are plotted as a function of $\gamma$ for $t / U = 0.12$ and $\Gamma / U = 0.0473$.
}
\label{fig:mhx}
\end{figure}
\begin{table}
\caption{
Relevant TTQD orbital states in the Kondo screening process, where $E_+$, $E_-$, and $A_-$
are the representations of wave functions in Eqs.~(\ref{eqn:phiE+1}), (\ref{eqn:phiE-1}), and
(\ref{eqn:phiA-}), respectively, and the orbital mixing depends on the ASO coupling $\gamma$.
The KIEP $\delta n^*$ exhibits an upturn at $\gamma= \gamma_{\rm min}$ ($\simeq 0.2$ in
Fig.~\ref{fig:nxtcr}).
There are three sequent $\gamma$ regions characterized by $\delta n^*$ and the local magnetic
susceptibility:
(i) Only the $E_+$ component participates in the Kondo screening process and gives each
maximum of $\delta n^*$ and $\chi = \chi_a + \chi_{b+c}$.
(ii) The $E_-$ component gives a negative contribution to $\delta n^*$, and the $b$-$c$
triplet state of $E_-$ is responsible for the diamagnetic behavior of $\chi_{b+c}$.
(iii) The mixing weight of $E_-$ is reduced by the $A_-$ components coupled to $E_+$ in the
Kondo singlet, leading to a gradual increase in $\delta n^*$ and decreases in $| \chi_a |$ and
$| \chi_{b+c} |$.
}
\label{t1}
\begin{center}
\begin{tabular}{ccc}
\hline
\multicolumn{1}{c}{} & \multicolumn{1}{c}{ASO coupling}
& \multicolumn{1}{c}{Relevant orbitals for Kondo screening} \\
\hline
(i) & $\gamma = 0$ & $E_+$ \\
(ii) & $0 < \gamma < \gamma_{\rm min}$ & $E_+ \oplus E_-$ \\
(iii) & $\gamma > \gamma_{\rm min}$ & $E_+ \oplus E_- \oplus A_-$ \\
\hline
\end{tabular}
\end{center}
\end{table}
The ASO coupling effect appears more significantly in both $\delta n^* (\gamma)$
and $T_{\rm cr} (\gamma)$ for a smaller $\Gamma / U$.
The most marked feature is the strong suppression of $\delta n^*$ by weak $E_\pm$ mixing for
a very small $\gamma$.
This indicates that $| \psi_{\sigma}^{E_-} \rangle$ in Eq.~(\ref{eqn:psiE}) also
participates in the Kondo singlet with $| \psi_{- \sigma}^{E_+} \rangle$ at low $T$.
More precisely, the spin-up conduction electrons couple with
$| \psi_\downarrow^{E_+} \rangle$ and $| \psi_\uparrow^{E_-} \rangle$, while the spin-down conduction electrons couple with
$| \psi_\uparrow^{E_+} \rangle$ and $| \psi_\downarrow^{E_-} \rangle$.
This is a consequence of the spin and orbital coupling classified by $\eta_z$ in Eq.~(\ref{eqn:eta}).
The $E_+$ state generates a positive electric polarization $\delta n^* > 0$ for $\gamma = 0$.
In Fig.~\ref{fig:nxtcr}, the mixing weight of $E_-$ in the ground state increases abruptly with
$\gamma$ ($< 0.1$), resulting in a negative contribution to $\delta n^*$.
On the other hand, the $A_-$ states in Eq.~(\ref{eqn:phiA-}) give no direct contribution to
$\delta n^*$ but effectively reduce the mixing weight of $E_-$ through the ASO coupling.
As this $A_-$ contribution becomes more relevant at $\gamma > 0.2$, $\delta n^*$ shows a
gradual increase with $\gamma$.
Thus, the entire $\gamma$ dependence of $\delta n^*$ is explained by the orbital symmetry
mixing of $E_\pm$ and $A_-$ in the ground state as summarized in Table~\ref{t1} (also see
Sect.~4.2).
\par

In Fig.~\ref{fig:nxtcr}, $T_{\rm cr}$ shows a similar $\gamma$ dependence to
$\delta n^*$, indicating that the emergence of the electric polarization is closely associated with
the Kondo effect.
As mentioned above, the even-parity state $| \psi_\sigma^{E_+} \rangle$ is only responsible for
the Kondo singlet with conduction electrons at $\gamma = 0$.
For a finite $\gamma$ ($< \gamma_{\rm min} \simeq 0.2$), this Kondo coupling is weakened
effectively by the ASO interaction since the odd-parity state $| \psi_{- \sigma}^{E_-} \rangle$,
which is not relevant to the Kondo effect, also participates in Kondo singlet formation and
reduces the mixing weight of $| \psi_\sigma^{E_+} \rangle$ in the ground state at low $T$ (see
Table~\ref{t1}).
In the $b$-$c$ bonding state at $T \rightarrow 0$, a spin singlet coexists with a spin triplet as
clearly shown in Fig.~\ref{fig:mhx}.
Here, we plot $\chi_a$ for the $a$-site magnetic susceptibility and $\chi_{b+c}$ for the sum
of the $b$- and $c$-site susceptibilities as a function of $\gamma$, in addition to the total
$\chi = \chi_a + \chi_{b+c}$.
For instance, let us consider that the spin "up" of the electron is parallel to the direction of the
applied magnetic field.
In this case, the dominant contributions to $\chi_a$ and $\chi_{b+c}$ are from
$| \psi_\uparrow^{E_+} \rangle$ and $| \psi_\downarrow^{E_-} \rangle$, respectively.
We find that the local magnetization is polarized in space, where the antiparallel local moments
are generated at one apex ($a$-site) and the opposite side ($b$-$c$ bond) of the triangular cluster.
On the other hand, the total $\chi$ monotonically decreases with increasing $\gamma$.
At $\gamma \simeq 0.15$, the diamagnetic $\chi_{b+c}$ exhibits an upturn and a gradual
increase with $\gamma$, indicating an increase in the mixing weights of
$| \phi_{3/2}^{A_-} \rangle$ and $| \phi_\downarrow^{A_-} \rangle$ in Eq.~(\ref{eqn:phiA-}), which
are coupled to $| \psi_\uparrow^{E_+} \rangle$ as well as $| \psi_\downarrow^{E_-} \rangle$ in the
ground state.

\subsection{Analogy with magnetic flux effect on orbital-parity mixing}
\begin{figure}
\begin{center}
\hspace*{0.7cm}
\includegraphics[width=5cm,clip]{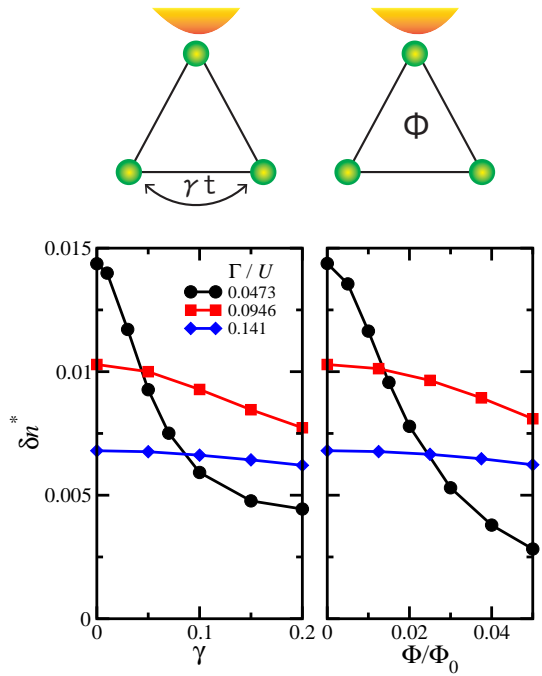}
\end{center}
\begin{center}
\includegraphics[width=7cm,clip]{fig5bottom.eps}
\end{center}
\caption{
(Color online)
Comparison of the ASO coupling effect (left panel) with the orbital effect of a weak magnetic field
(right panel) on $\delta n^*$ as a function of $\gamma$ and the magnetic flux $\Phi / \Phi_0$,
respectively ($\Phi_0$ is the magnetic flux quantum).
The data are for various $\Gamma / U$ values, where $t / U = 0.12$ is fixed.
}
\label{fig:nx}
\end{figure}
For a small ASO coupling $\gamma$, the abrupt decrease in $\delta n^*$ for a small
$\Gamma / U$ in Fig.~\ref{fig:nxtcr} is well explained by the orbital-parity mixing approximately
represented by $c_+ | \psi_\sigma^{E_+} \rangle + c_- | \psi_{- \sigma}^{E_-} \rangle$, where
$c_\pm$ is the mixing weight of the $E_\pm$ state.
It is confirmed that the ratio $| c_- | / | c_+ |$ considerably increases with $\gamma$ since
$\delta n^*$ is proportional to $| c_+ |^2 - | c_- |^2$.
The KIEP $\delta n^*$ is also sensitively dependent on
$\Gamma / U$ as shown in Fig.~\ref{fig:nx} (left).
The increase in $\Gamma / U$ causes the reduction of $\delta n^*$ at $\gamma = 0$, and
the charge distribution becomes uniform over the three sites of the triangular cluster.
This indicates that the $E_+$ state is more dominant in the ground state, namely,
$| c_- | / | c_+ | \simeq 0$ for a large Kondo coupling $\Gamma / U$.
In addition, the ASO coupling effect becomes irrelevant for the emergent electric polarization even
when $\gamma$ is larger.
\par

For the KIEP, there is a close analogy between the above ASO
coupling effect on the $b$-$c$ bond and the magnetic flux effect on the orbital states in
the triangular loop.
Both effects give rise to the mixing of the $E_\pm$ orbital parities and lead to a similar
dependence of $\delta n^*$ on the orbital-mixing parameters:
$\gamma$ for the ASO coupling and $\Phi / \Phi_0$ for the magnetic flux penetrating
through the triangle in Fig.~\ref{fig:nx}.
The data of $\delta n^*$ are plotted as a function of each parameter for comparison.
Appendix~C shows the magnetic flux contribution to the wave function of the lowest-energy state
for $\Gamma = 0$.
When the Kondo singlet is realized at low temperatures for a finite $\Gamma$, the $E_\pm$ orbital
mixing for a finite $\Phi$ is expressed approximately as
$c_+ | \psi_\sigma^{E_+} \rangle + c_- i | \psi_\sigma^{E_-} \rangle$ with the conservation of the
total spin $S$.
\cite{Koga14}
On the other hand, the ASO coupling effect generates spin-up and spin-down coupling as
$c_+ | \psi_\sigma^{E_+} \rangle + c_- | \psi_{- \sigma}^{E_-} \rangle$
in addition to the $E_\pm$ mixing, which maintains the conservation of $\eta_z = +1$ or
$\eta_z = -1$.
Thus, the ASO coupling $\gamma$ and magnetic flux $\Phi$ effects qualitatively give the same
contribution to $\delta n^*$ as internal and external fields, respectively, in spite of the
different orbital-parity mixing mechanisms.

\section{Summary and Concluding Remarks}
We have studied the local ASO coupling effect on the emergent electric polarization in a triangular
cluster of magnetic impurities at low temperatures, which is relevant in the absence of the inversion
symmetry.
One of the three local spins in the cluster ($\bS_a$) is coupled to the lead electrons, and the ASO
interaction introduced in the other coupled ($b$-$c$ bond) spins causes mixing of the
even and odd parities (symmetric and antisymmetric bonding) of orbital states ($E_\pm$ mixing).
This effect competes with the Kondo effect and gives rise to a considerable reduction of the
KIEP $\delta n^*$ at low temperatures for a small Kondo coupling
strength $\Gamma / U$.
It was also found that the ASO coupling dependence of $\delta n^*$ is strongly correlated with that
of the crossover temperature $T_{\rm cr}$.
The latter is an energy scale derived from a universal
function of the low-temperature dependence of the local magnetic susceptibility.
Another important result is that the ASO-coupled spins exhibit diamagnetism owing to an
equal-spin pair in the odd-parity $E_-$ state that participates in Kondo singlet formation
as well as the even-parity $E_+$ state.
We have also elucidated a close analogy between the ASO coupling and magnetic flux effects on
the orbital-parity mixing as internal and external field controls of $\delta n^*$, respectively.
\par

Here, the ASO coupling was introduced in one bond ($b$-$c$ bond) of the TTQD on
the opposite side of an apex connected to a lead.
This side effect was considered as a kind of boundary condition for moderate control of the
effective Kondo coupling with the lead electrons.
Indeed, the ASO coupling strength is a controllable parameter related to the mixing weight of
different-parity molecular orbitals.
For instance, this partially originates from a hybridization between intradot orbitals with different
parities such as the last term including $V_z$ in Eq.~(\ref{eqn:Hs+p}), which can be induced by a
local electric field in the absence of the inversion symmetry.
Experimentally, the local ASO coupling can be tuned by the gate voltage and acts as a
controllable boundary condition of such a Kondo system with a nanostructure.
\par

In Sect.~2, it was pointed out that the ASO-coupled spins generate a vector spin chirality,
which is closely associated with the parity mixing of the molecular orbital symmetries
(here, $E_+ \oplus E_-$ and $E_+ \oplus A_-$) accompanied by spin modulation.
Recently, we have reported a group-theoretical study of various spin-dependent electric dipoles
in the absence of the inversion symmetry, including the relationship between the vector spin
chirality of the two-site spins and possible electric dipoles induced by a magnetic field or magnetic
ordering.
\cite{Matsumoto17}
As another magnetoelectric effect, it will also be intriguing to reveal the interplay between the
KIEP $\delta n^*$ and the vector spin chirality generated by the ASO-coupled spins in the present
model.
A detailed analysis is left for a future study.
\par

We would also like to mention a symmetry-lowering effect on $\delta n^*$, which depends on
different intersite electron-hopping parameters $t_{ij}$ ($i, j = a, b, c$; $i \ne j$).
For the parity-symmetric case $t_{ab} = t_{ca}$ (fixed at $t$) in the absence of the ASO
coupling, $| \phi_{\sigma, 1}^{E_+} \rangle$ in Eq.~(\ref{eqn:phiE+1}) is stabilized by the Kondo
effect, while $| \phi_{\sigma, 1}^{E_-} \rangle$ in Eq.~(\ref{eqn:phiE-1}) is favored by lowering
$t_{bc}$ from $t$.
This competition leads to a quantum transition from the former to the latter upon lowering $t_{bc}$.
The critical point $t_{\rm c} \lesssim t$ is obtained for a sufficiently small $\Gamma / U$.
This quantum transition was investigated in several theoretical studies as a unique feature of the
Kondo effect for a triangular loop of three magnetic impurities.
\cite{Mitchell09,Mitchell13,Tooski14}
In the present study, we have focused on the KIEP in the equilateral TTQD, and the result also
holds for the lower-symmetry case $t_{bc} > t_{\rm c}$.
\par

The recent development of a scanning tunneling microscope (STM) experiment has realized the
manipulation of atoms and their precise arrangement in designed configurations on a surface
template.
Using such a fabrication technique, it is now possible to create coupled QDs with a perfect
geometric structure.
\cite{Folsch14}
The STM image visualizes the charge distributions associated with the geometry, and the parity
symmetry of a wave function is reflected in the electron density maps.
In addition, STM spectroscopy unveils quantum spin configurations in artificially coupled spin
chains at the atomic scale.
\cite{Heinrich04,Hirjibehedin06,Wiesendanger09}
Thus, it is very promising for detecting a nanoscale electric polarization correlated with local spin
states, and the confirmation of the KIEP will stimulate a new
application of the Kondo effect to spintronic and multiferroic devices.
From a different viewpoint of the experiments, the ASO-coupling-dependent KIEP is expected to
give a relevant contribution to the linear conductance between two leads through a single QD of
the TTQD, which can be mapped to the present Anderson model for the TTQD with a single lead.
Indeed, for a weak dot-lead contact, the ASO coupling effect causes a considerable reduction of
the crossover temperature $T_{\rm cr}$ in Fig.~\ref{fig:nxtcr}, which corresponds to the Kondo
temperature.
As shown in Sect.~4.2, the ASO coupling can be considered as a magnetic flux through the TTQD
and could be chosen as a controllable parameter for detecting an interference effect on the
conductance, as proposed in several theoretical studies on QD systems with various
configurations.
\cite{Izumida97,Kuzmenko06,Akera93,Bruder96,Hofstetter01,Tanaka06,Komijani13}

\acknowledgments
This work was supported by JSPS KAKENHI Grant Numbers
16H01070 (J-Physics), 15H05885 (J-Physics), 26400332, and 15K05176.

\appendix
\section{Effective Spin-Orbit Interaction Due to Orbital-Parity Mixing}
The ASO interaction plays an important role in electron transfer between two QDs.
Here, we show a simple two-site model with different intradot orbital-parity mixing.
For simplicity, we consider that even- and odd-parity orbitals are represented by, for instance,
$s$- and $p$-orbitals, respectively, and the latter orbitals are coupled to spins through the SO
interaction at each site as realized in real atoms.
This atomlike assumption may be reasonable for local electrons strongly confined in a QD
by a sort of central field, although the intradot SO coupling is extremely weak as usual.
For the $p$-orbital wave functions, the three components
$| p_x \rangle$, $| p_y \rangle$, and $| p_z \rangle$ are related to the $l = 1$ angular momentum
states as
\begin{align}
& l_z = 1:~~ | p_+ \rangle = \frac{1}{\sqrt{2}} ( - | p_x \rangle - i | p_y \rangle ),
\label{eqn:lz1} \\
& l_z = 0:~~ | p_z \rangle, \\
& l_z = - 1:~~ | p_- \rangle = \frac{1}{\sqrt{2}} ( | p_x \rangle - i | p_y \rangle ).
\label{eqn:lz-1}
\end{align}
We assume that the $p$-orbital degeneracy is lifted by an axial crystal field and that
the $p_z$ orbital energy is much lower than that of $p_x$ and $p_y$.
The local $p$-orbital state with spin is described by the onsite Hamiltonian
\begin{align}
& H_p = H_{\rm CEF} + H_{\rm LS}
\nonumber \\
&~~~~
 = \Delta ( n_{p_x} + n_{p_y} ) + \lambda \bl \cdot \bs,
\end{align}
where $\Delta$ ($>0$) is the crystal-field energy measured from the $p_z$ orbital energy level
and $n_{p_x}$ ($n_{p_y}$) is the number of $p_x$ ($p_y$) electrons in the first term.
In the second term, $\lambda$ ($>0$) represents the SO coupling constant and
\begin{align}
\bl \cdot \bs = \frac{1}{2} l_+  s_- + \frac{1}{2} l_- s_+ + l_z s_z,
\end{align}
where for $l_\pm$ ($\equiv l_x \pm i l_y$),  the finite matrix elements of $( l_\pm )_{mm'}$ and
$( l_z )_{mm'}$ ($m,m' = 0, \pm 1$ for $l_z$) are given by
\begin{align}
& ( l_+ )_{1,0} = ( l_+ )_{0,-1} = ( l_- )_{0,1} = ( l_- )_{-1,0} = \sqrt{2}, \\
& ( l_z )_{1, 1} = 1,~~( l_z )_{-1,-1} = - 1,
\end{align}
and $\bs$ ($\equiv \bsig / 2$) is a spin operator ($s_\pm \equiv s_x \pm i s_y$).
For a small SO coupling $\lambda \ll \Delta$, the lowest-energy eigenstates of $H_p$ are
obtained as
\begin{align}
& | P, + \rangle = | p_{z, \uparrow} \rangle
 - \frac{\lambda}{\sqrt{2} \Delta} | p_{+,\downarrow} \rangle,
\nonumber \\
& | P, - \rangle = | p_{z, \downarrow} \rangle
 - \frac{\lambda}{\sqrt{2} \Delta} | p_{-,\uparrow} \rangle,
\end{align}
with the eigenenergy $- \lambda^2 / (2 \Delta)$.
\par

Next, we introduce the even- and odd-parity mixing on the basis of the $s$-$p$ hybridization
using the following Hamiltonian:
\cite{Yanase11}
\begin{align}
& H_{s+p} = H_p + H_s + H_{s, p_z}
\nonumber \\
&~~~~~~
 = H_p - E_s n_s + V_z \sum_\sigma ( | s, \sigma \rangle \langle p_z, \sigma | + {\rm h. c.} ),
\label{eqn:Hs+p}
\end{align}
where the second term with $- E_s$ ($< 0$) represents the energy of the $s$-electrons with the
number operator $n_s$, and $V_z$ in the last term represents the deviation from the inversion
symmetry, namely, the hybridization strength between $s$- and $p_z$-orbitals.
For a small $| V_z |$ ($ \ll E_s $), the lowest-energy states of $H_{s+p}$ are dominated by the
$s$-electron, and the wave functions are obtained as
\begin{align}
& | S, + \rangle = | s, \uparrow \rangle - \frac{V_z}{E_s} | P, + \rangle
\nonumber \\
&~~~~~~
 = | s, \uparrow \rangle -  \frac{V_z}{E_s} | p_z, \uparrow \rangle
 + \frac{V_z \lambda}{\sqrt{2} E_s \Delta} | p_+, \downarrow \rangle,
\nonumber \\
& | S, - \rangle = | s, \downarrow \rangle - \frac{V_z}{E_s} | P, - \rangle
\nonumber \\
&~~~~~~
 = | s, \downarrow \rangle -  \frac{V_z}{E_s} | p_z, \downarrow \rangle
 + \frac{V_z \lambda}{\sqrt{2} E_s \Delta} | p_-, \uparrow \rangle.
\label{eqn:S+-}
\end{align}
These spin and orbitally coupled states are used for the derivation of an intersite electron
hopping-type Hamiltonian.
For simplicity, we assume that two sites (labeled 1 and 2) are aligned in the $x$ direction.
The position of site 1 (2) is on the left (right) side.
When the electron creation and annihilation at the $i$th site ($i = 1, 2$) are represented by
$c_{i, \mu, \sigma}^\dagger$ and $c_{i, \mu, \sigma}$, respectively, for the $\mu$ ($= s, p_+, p_-$)
orbitals with spin $\sigma$, the intersite orbital mixing is described by
\begin{align}
& H_{\rm mix} = - t_{ss} \sum_\sigma c_{1, s, \sigma}^\dagger c_{2, s, \sigma}
\nonumber \\
&~~~~~~~~~~
 - t_{sp} ( - c_{1, s, \uparrow}^\dagger c_{2, p_- \uparrow}
 + c_{1, p_-, \uparrow}^\dagger c_{2, s, \uparrow} )
\nonumber \\
&~~~~~~~~~~
 -t_{sp} ( c_{1, s, \downarrow}^\dagger c_{2, p_+ \downarrow}
 - c_{1, p_+, \downarrow}^\dagger c_{2, s, \downarrow} ) + {\rm h. c.},
\label{eqn:Hmix}
\end{align}
where $t_{ss}$ ($t_{sp}$) is the transition matrix element between the two-site $s$-orbitals
(between the $s$-orbital on one site and the $p_x$-orbital on the other).
We have discarded the $p$-$p$ electron-hopping term since its contribution is considered to be
much smaller than those of the $s$-$s$ and $s$-$p$ terms.
The phase difference $\pi$ arises in the $s \rightarrow p_x$ and $p_x \rightarrow s$ transfers
between site 1 and site 2 owing to the odd parity of the $p$-orbital.
On the basis of Eq.~(\ref{eqn:Hmix}), we calculate the matrix elements of the overlap integrals
$\langle i, S, \pm | j, S, \pm \rangle$ and $\langle i, S, \pm | j, S, \mp \rangle$
($i,j = 1,2$ are the site indices and $i \ne j$) in Eq.~(\ref{eqn:S+-}).
Replacing the notations $\{ +, - \}$ by $\{ \uparrow, \downarrow \}$ respectively, we finally
obtain the electron-hopping Hamiltonian
\begin{align}
H_{ss} = - t_{ss} \sum_\sigma ( c_{1, S, \sigma}^\dagger c_{2, S, \sigma} + {\rm h. c.})
\end{align}
between same-parity orbitals ($s$-orbitals) and the local ASO interaction Hamiltonian
\begin{align}
H_{sp} = \gamma t_{ss} ( c_{1, S, \uparrow}^\dagger c_{2, S, \downarrow}
 - c_{1, S, \downarrow}^\dagger c_{2, S, \uparrow} ) + {\rm h. c.}
\label{eqn:Hsp}
\end{align}
for different-orbital-parity mixing ($s$-$p$ orbital mixing), where the ASO coupling constant is
given by
\begin{align}
\gamma = \sqrt{2} \frac{t_{sp}}{t_{ss}} \frac{V_z}{E_s} \frac{\lambda}{\Delta}
\label{eqn:gamma}
\end{align}
in the unit of $t_{ss}$.
As a consequence, in terms of the $s$-dominant orbital basis, the $s$-$p_x$ electron transfer is
transformed to the ASO interaction.
Equation~(\ref{eqn:Hso1}) is obtained by the substitutions $\gamma t_{ss} \rightarrow \lambda_A$
and $c_{i, S, \sigma}^\dagger$ ($c_{i, S, \sigma}$) $\rightarrow$
$d_{i \sigma}^\dagger$ ($d_{i \sigma}$) in Eq.~(\ref{eqn:Hsp}).

\section{Three-Site Hubbard Model for TTQD}
Here, we consider the half-filled case of the triangular spin cluster for the strong-$U$ limit using the
following three-site Hubbard model Hamiltonian:
\begin{align}
& H_{\rm trimer} = H_U + H_t
\nonumber \\
&~~~~~~
 = U \sum_i n_{i \uparrow} n_{i \downarrow}
  - t \sum_\sigma \sum_{i \ne j} d_{i \sigma}^\dagger d_{j \sigma},
\label{eqn:Htri}
\end{align}
where $i$ and $j$ ($= a, b, c$) denote the three identical sites with the same electron-hopping
matrix element $t$ in $H_t$.
To obtain the lowest-energy states of $H_{\rm trimer}$, we start from the $t = 0$ case, and next
treat a finite-$t$ effect as a perturbation.
Owing to the onsite Coulomb interaction term $H_U$, a single electron is occupied at each site for
the lowest-energy states.
One of the lowest $S_z = \pm 1/2$ states with $S = 1/2$ is given by
\begin{align}
| \phi_{\sigma,1}^{E_+} \rangle
 = \frac{1}{\sqrt{2}} d_{a,\sigma}^\dagger
 ( d_{b,\sigma}^\dagger d_{c, - \sigma}^\dagger
  - d_{b, - \sigma}^\dagger d_{c,\sigma}^\dagger ) | 0 \rangle,
\label{eqn:phiE+1}
\end{align}
where $E_+$ represents an even-parity representation of the degenerate orbitals with the $C_3$
symmetry
\cite{Janani14}
and the subscript $\sigma$ of $| \phi_{\sigma, 1}^{E_+} \rangle$ represents $S_z$.
On the right-hand side, $- \sigma$ ($= \downarrow, \uparrow$) is the time-reversal component of
$\sigma$ ($= \uparrow, \downarrow$), respectively.
In the same manner, the other orbitally degenerate $S_z = \pm 1/2$ state with an odd parity $E_-$
is written as
\begin{align}
& | \phi_{\sigma,1}^{E_-} \rangle
 = \frac{1}{\sqrt{6}}
 [ 2 d_{a, - \sigma}^\dagger d_{b,\sigma}^\dagger d_{c,\sigma}^\dagger
\nonumber \\
&~~~~~~~~~~~~~~~~~~
  - d_{a,\sigma} ( d_{b,\sigma}^\dagger d_{c, - \sigma}^\dagger
  + d_{b, - \sigma}^\dagger d_{c,\sigma}^\dagger )] | 0 \rangle.
\label{eqn:phiE-1}
\end{align}
In addition, the $S = 3/2$ states
\begin{align}
& | \phi_{3/2}^{A_-} \rangle
 =  d_{a \uparrow}^\dagger d_{b \uparrow}^\dagger d_{c \uparrow}^\dagger | 0 \rangle,~~
 | \phi_{- 3/2}^{A_-} \rangle
 =  d_{a \downarrow}^\dagger d_{b \downarrow}^\dagger d_{c \downarrow}^\dagger | 0 \rangle,~~ 
\nonumber \\
& | \phi_\sigma^{A_-} \rangle
 =  \frac{1}{\sqrt{3}}
  ( d_{a,\sigma}^\dagger d_{b,\sigma}^\dagger d_{c, - \sigma}^\dagger
   +  d_{a,\sigma}^\dagger d_{b, - \sigma}^\dagger d_{c,\sigma}^\dagger
 \nonumber \\
&~~~~~~~~~~~~~~~~~~
   +  d_{a, - \sigma}^\dagger d_{b,\sigma}^\dagger d_{c,\sigma}^\dagger ) | 0 \rangle
\label{eqn:phiA-}
\end{align}
are categorized into a one-dimensional representation $A_-$ with an odd party.
We note that the $E_+$ ($E_-$, $A_-$) wave function is even (odd) with respect to the interchange
of the site indices $b \leftrightarrow c$.
\par

Through the intersite electron hopping, $| \phi_{\sigma,1}^{E_+} \rangle$ is
coupled to the doubly occupied states with an excitation energy of $U$ as
\begin{align}
\left\{
\begin{array}{l}
| \phi_{\sigma,2}^{E_+} \rangle = \ds{\frac{1}{\sqrt{2}}} d_{a,\sigma}^\dagger
 ( d_{b,\sigma}^\dagger d_{b, - \sigma}^\dagger + d_{c,\sigma}^\dagger d_{c, - \sigma}^\dagger )
 | 0 \rangle, \\
| \phi_{\sigma,3}^{E_+} \rangle = \ds{\frac{1}{\sqrt{2}}}
 d_{a,\sigma}^\dagger d_{a, - \sigma}^\dagger
 ( d_{b,\sigma}^\dagger + d_{c,\sigma}^\dagger ) | 0 \rangle, \\
| \phi_{\sigma,4}^{E_+} \rangle = \ds{\frac{1}{\sqrt{2}}}
 ( d_{b,\sigma}^\dagger d_{b, - \sigma}^\dagger d_{c,\sigma}^\dagger
  + d_{b,\sigma}^\dagger d_{c,\sigma}^\dagger d_{c, - \sigma}^\dagger ) | 0 \rangle.
\end{array}
\right.
\end{align}
In the same manner, $| \phi_{\sigma,1}^{E_-} \rangle$ is coupled to
\begin{align}
\left\{
\begin{array}{l}
| \phi_{\sigma,2}^{E_-} \rangle = \ds{\frac{1}{\sqrt{2}}} d_{a,\sigma}^\dagger
 ( d_{b,\sigma}^\dagger d_{b, - \sigma}^\dagger - d_{c,\sigma}^\dagger d_{c, - \sigma}^\dagger )
 | 0 \rangle, \\
| \phi_{\sigma,3}^{E_-} \rangle = \ds{\frac{1}{\sqrt{2}}}
 d_{a,\sigma}^\dagger d_{a, - \sigma}^\dagger
 ( d_{b,\sigma}^\dagger - d_{c,\sigma}^\dagger ) | 0 \rangle, \\
| \phi_{\sigma,4}^{E_-} \rangle = \ds{\frac{1}{\sqrt{2}}}
 ( - d_{b,\sigma}^\dagger d_{b, - \sigma}^\dagger d_{c,\sigma}^\dagger
  + d_{b,\sigma}^\dagger d_{c,\sigma}^\dagger d_{c, - \sigma}^\dagger ) | 0 \rangle.
\end{array}
\right.
\end{align}
The $4 \times 4$ matrix form of
$\langle \phi_{\sigma,i}^{E_+} | H_t | \phi_{\sigma,j}^{E_+} \rangle$ ($i,j = 1, 2, 3, 4$) is obtained as
\begin{align}
t
\left(
\begin{array}{cccc}
0 & -2 & 1 & 1 \\
-2 & 0 & 1 & -1 \\
1 & 1 & -1 & 0 \\
1 & -1 & 0 & 1
\end{array}
\right),
\label{eqn:HtE+}
\end{align}
and for the $E_-$ states, the $H_t$ matrix is given by
\begin{align}
t
\left(
\begin{array}{cccc}
0 & 0 & - \sqrt{3} & - \sqrt{3} \\
0 & 0 & 1 & 1 \\
- \sqrt{3} & 1 & 1 & 0 \\
- \sqrt{3} & 1 & 0 & -1
\end{array}
\right).
\label{eqn:HtE-}
\end{align}
Let us apply the perturbation of $H_t$ to the unperturbed states
$| \phi_{\sigma,1}^{E_+} \rangle$ and $| \phi_{\sigma,1}^{E_-} \rangle$ of $H_U$.
Up to the second order of $\bar{t}$ ($\equiv t / U$), it leads to the following lowest-lying states with
the fourfold degeneracy ($E_{\pm}$ and $\sigma = \uparrow, \downarrow$):
\cite{Bulaevskii08}
\begin{align}
& | \psi_\sigma^{E_+} \rangle
 = (1 - 3 \bar{t}^2) | \phi_{\sigma,1}^{E_+} \rangle + 2 \bar{t} | \phi_{\sigma,2}^{E_+} \rangle
\nonumber \\
&~~~~~~
  + (- \bar{t} - 3 \bar{t}^2) | \phi_{\sigma,3}^{E_+} \rangle
  + (- \bar{t} + 3 \bar{t}^2) | \phi_{\sigma,4}^{E_+} \rangle,
\nonumber \\
& | \psi_\sigma^{E_-} \rangle
 = (1 - 3 \bar{t}^2) | \phi_{\sigma,1}^{E_-} \rangle - 2 \sqrt{3} \bar{t}^2 | \phi_{\sigma,2}^{E_-} \rangle
\nonumber \\
&~~~~~~
  + \sqrt{3} (\bar{t} - \bar{t}^2) | \phi_{\sigma,3}^{E_-} \rangle
  + \sqrt{3} (\bar{t} + \bar{t}^2) | \phi_{\sigma,4}^{E_-} \rangle,
\label{eqn:psiE}
\end{align}
with an energy of $- 6 t^2 / U$.
\par

In Eq.~(\ref{eqn:Hd}), the ASO interaction is introduced in one of the three bonds,
i.e., the $b$-$c$ bond in the triangle cluster, and the corresponding Hamiltonian is given by
replacing $(1,2)$ with $(b,c)$ in Eq.~(\ref{eqn:Hso1}).
Since the $a$-site orbital has an even parity ($\tau_z = +1$), the spin-up
($d_{a,\uparrow}^\dagger$) and spin-down ($d_{a,\downarrow}^\dagger$) electron states are
categorized into $\tau_z \sigma_z = +1$ and $\tau_z \sigma_z = -1$, respectively.
Here, we choose the half-filled states of the $\eta_z = -1$ type
$| \psi_\uparrow^{E_+} \rangle$, $| \psi_\downarrow^{E_-} \rangle$,
$| \phi_{3/2}^{A_-} \rangle$, and $| \phi_\downarrow^{A_-} \rangle$:
the corresponding $S_z$ values of the total spins are $1/2$, $- 1/2$, $3/2$, and $-1/2$,
respectively.
These four states are coupled to each other through $H_{\rm so}(b,c)$ as
\begin{align}
& \langle \psi_\downarrow^{E_-} | H_{\rm so}(b,c) | \psi_\uparrow^{E_+} \rangle
 = - \frac{1}{\sqrt{3}} \xi,
\nonumber \\
& \langle \phi_{3/2}^{A_-} | H_{\rm so}(b,c) | \psi_\uparrow^{E_+} \rangle
 = - \frac{1}{\sqrt{2}} \xi,
\nonumber \\
& \langle \phi_\downarrow^{A_-} | H_{\rm so}(b,c) | \psi_\uparrow^{E_+} \rangle
 = - \frac{1}{\sqrt{6}} \xi,
\label{eqn:Hso-bc}
\end{align}
where $\xi = 4 \gamma \bar{t} t$ and higher-order terms of $\bar{t}$ are neglected.
\par

Here, we show an analogy with the vector spin chirality in the two-site case described in Sect.~2.
Let us assume that $| \psi_\uparrow^{E_+} \rangle$ has the lowest energy and the other three 
excited states are almost degenerate, keeping in mind that the $E_+$ state is stabilized by
the Kondo effect in the TTQD.
For $| \psi_\uparrow^{E_+} \rangle$, the perturbation of $H_{\rm so}$ in Eq.~(\ref{eqn:Hso-bc})
gives the lowest-lying state as
\begin{align}
& \alpha \left. \left| \psi_\uparrow^{E_+} \right. \right\rangle
 + \beta \left( \frac{1}{\sqrt{3}} \left. \left| \psi_\downarrow^{E_-} \right. \right\rangle \right.
\nonumber \\
&~~~~~~~~~~~~~~~~~~\left.
  + \frac{1}{\sqrt{2}} \left. \left| \phi_{3/2}^{A_-} \right. \right\rangle
  + \frac{1}{\sqrt{6}} \left. \left| \phi_\downarrow^{A_-} \right. \right\rangle \right)
\nonumber \\
&~~~~~~
\simeq \frac{1}{\sqrt{2}} d_{a \uparrow}^\dagger
 \left[ \alpha ( d_{b \uparrow}^\dagger d_{c \downarrow}^\dagger
   - d_{b \downarrow}^\dagger d_{c \uparrow}^\dagger ) \right.
\nonumber \\
&~~~~~~~~~~~~~~~~~~\left.
  + \beta ( d_{b \uparrow}^\dagger d_{c \uparrow}^\dagger
   + d_{b \downarrow}^\dagger d_{c \downarrow}^\dagger ) \right] | 0 \rangle,
\label{eqn:psi4}
\end{align}
where $\alpha$ and $\beta$ are real coefficients that satisfy $\alpha^2 + \beta^2 = 1$.
The ratio of the matrix elements in Eq.~(\ref{eqn:Hso-bc}) is directly reflected in the coefficients
of $| \psi_\downarrow^{E_-} \rangle$, $| \phi_{3/2}^{A_-} \rangle$, and
$| \phi_\downarrow^{A_-} \rangle$.
One can find that the vector spin chirality $(\bS_b \times \bS_c)_y$ is induced by the mixing of the
$b$-$c$ singlet and triplet states in Eq.~(\ref{eqn:psi4}), corresponding to $| \phi_S \rangle$ and
$-i | \phi_{T_y} \rangle$, respectively,  in Eq.~(\ref{eqn:psiSDT}) in the two-site case.

\section{$E_\pm$ Parity Mixing by Magnetic Flux through TTQD}
As well as the ASO interaction, a magnetic flux effect causes the $E_\pm$ orbital-parity mixing
between $| \phi_{\sigma,i}^{E_+} \rangle$ and $| \phi_{\sigma,j}^{E_-} \rangle$ given in
Appendix~B.
\cite{Wang08}
In this case, the total spin $S$ is conserved.
For the equilateral triangle with $C_3$ symmetry, the electron hopping matrix element of
$H_t$ in Eq.~(\ref{eqn:Htri}) is modulated by the magnetic flux $\Phi$ penetrating through the
triangular loop as
\begin{align}
H_{t, \Phi} = - \sum_\sigma \sum_{i \ne j} t e^{ i \varphi_{ij} } d_{i \sigma}^\dagger d_{j \sigma},
\end{align}
where $\varphi_{ab} = \varphi_{bc} = \varphi_{ca} = \varphi / 3$ ($\varphi_{ij} = - \varphi_{ji}$) and
the phase $\varphi = 2 \pi \Phi / \Phi_0$ is defined by the magnetic flux quantum $\Phi_0 = hc/e$
($h$: Planck constant, $c$: speed of light in vacuum, $e$: elementary charge).
The matrix elements of $\langle \phi_{\sigma,i}^{E_+} | H_{t, \Phi} | \phi_{\sigma,j}^{E_+} \rangle$
and $\langle \phi_{\sigma,i}^{E_-} | H_{t, \Phi} | \phi_{\sigma,j}^{E_-} \rangle$ ($i,j = 1, 2, 3, 4$) are
given by replacing $t$ with $t \cos (\varphi / 3)$ in Eqs.~(\ref{eqn:HtE+}) and (\ref{eqn:HtE-}),
respectively.
For the $E_\pm$ mixing, the $4 \times 4$ matrix form of
$\langle \phi_{\sigma,i}^{E_+} | H_{t, \Phi} | \phi_{\sigma,j}^{E_-} \rangle$ is
obtained as
\begin{align}
- i t \sin \frac{\varphi}{3}
\left(
\begin{array}{cccc}
0 & -2 & -1 & 1 \\
0 & 0 & 1 & 1 \\
-\sqrt{3} & -1 & -1 & 0 \\
\sqrt{3} & 1 & 0 & -1
\end{array}
\right),
\end{align}
and the $E_\pm$ orbital degeneracy is lifted.
For a small $| \Phi |$, the off-diagonal matrix element of $H_{t, \Phi}$ is given by
\begin{align}
\langle \psi_\sigma^{E_+} | H_{t, \Phi} | \psi_\sigma^{E_-} \rangle
\simeq - i 6 \sqrt{3} \bar{t}^2 t \varphi
\end{align}
on the basis of Eq.~(\ref{eqn:psiE}).
Thus, the orbital effect of the magnetic field leads to the lowest-energy eigenstates as
$| \psi_\sigma^{E_+} \rangle \pm i | \psi_\sigma^{E_-} \rangle$ for the three-site Hubbard model.


\end{document}